\documentclass[12pt]{article}
\makeindex

\def\keywords#1{\small\centerline{\bf Key Words}\vspace{5mm}\centerline{\parbox{14cm}{#1}}}

\def\gtapprox{\buildrel{\lower.7ex\hbox{$>$}}\over
                       {\lower.7ex\hbox{$\sim$}}}
\def\nq{\hspace{-1em}}

\def\ignore#1{}

\def\odt{{\textstyle{1\over 2}}}

\def\hbar{h\!\!\!\!^{-}\,}

\def\eps{\varepsilon}
\def\beq{\begin{equation}}
\def\eeq{\end{equation}}
\def\beqn{\begin{displaymath}}
\def\eeqn{\end{displaymath}}
\def\bqa{\begin{equation}\begin{array}{c}}
\def\eqa{\end{array}\end{equation}}
\def\bqan{\begin{displaymath}\begin{array}{c}}
\def\eqan{\end{array}\end{displaymath}}

\def\length{{l}}

\begin{document}


\begin{titlepage}

\begin{center}  \vspace*{2cm}
  {\small Technical Report IDSIA-16-00, 7. December 2000}\\[2cm]
  {\huge\sc An effective Procedure for} \\[0.5cm]
  {\huge\sc Speeding up Algorithms}      \\[1cm]
  {\bf Marcus Hutter}                    \\[1cm]
  {\rm IDSIA, Galleria 2, CH-6928 Manno-Lugano, Switzerland}  \\
  {\rm\footnotesize marcus@idsia.ch \qquad http://www.idsia.ch} \\[1cm]
\end{center}



\hfill

\begin{abstract}
The provably asymptotically fastest algorithm within a factor of 5
for formally described problems will be constructed. The main idea
is to enumerate all programs provably equivalent to the original
problem by enumerating all proofs. The algorithm could be
interpreted as a generalization and improvement of Levin search,
which is, within a multiplicative constant, the fastest algorithm
for inverting functions. Blum's speed-up theorem is avoided by
taking into account only programs for which a correctness proof
exists. Furthermore, it is shown that the fastest program that
computes a certain function is also one of the shortest programs
provably computing this function. To quantify this statement, the
definition of Kolmogorov complexity is extended, and two new
natural measures for the complexity of a function are defined.
\end{abstract}

\end{titlepage}

\section{Introduction}\label{secInt}
Searching for fast algorithms to solve certain problems is a central
and difficult task in computer science.
Positive results usually come from explicit constructions of
efficient algorithms for specific problem classes. Levin's
algorithm is one of the few general purpose speed-ups. Within a
(large) factor, it is the fastest algorithm to invert a function
$g\!:\!Y\to\!X$, if $g$ is easy to evaluate
\cite{Levin:73,Levin:84}. Given $x$, an inversion algorithm $p$
tries to find a $y$ with $g(y)\!=\!x$ by evaluating $g$ on a trial
sequence $y_i\!\in\!Y$. Levin search runs all such algorithms $p$
in parallel with relative computation time $2^{-\length(p)}$;
i.e.\ a time fraction $2^{-\length(p)}$ is devoted to execute $p$,
where $l(p)$ is the length of program $p$ (coded binary). The
total computation time to find a solution (if one exists) is
bounded by $2^{\length(p)}\!\cdot\!time_{p}$, where $p$ is any
program of length $\length(p)$ finding a solution in time
$time_{p}$. Hence, Levin search is optimal within a multiplicative
constant in computation time. It can be modified to handle
time-limited optimization problems as well \cite{Solomonoff:86}.
Many, but not all problems are of inversion or optimization type.
The matrix multiplication example, considered in the next section,
for instance, cannot be brought into this form. Furthermore, the
large factor $2^{\length(p)}$ somewhat limits the applicability.

A wider class of problems can be phrased in the following way.
Given a formal specification of a problem depending on some
parameter $x\!\in\!X$, we are interested in a fast algorithm
computing solution $y\!\in\!Y$. This means that we are
interested in a fast algorithm computing $f\!:\!X\!\to\!Y$, where
$f$ is a formal specification of the problem. For function
inversion problems, $f\!:=\!g^{-1}$. Ideally, we would like to
have the fastest algorithm, maybe apart from some small constant
factor in computation time. Unfortunately, Blum's Speed-up Theorem
\cite{Blum:67,Blum:71} shows that there are problems for which an
(incomputable) sequence of speed-improving algorithms (of
increasing size) exists, but no fastest algorithm, however.

In the approach presented here, we consider only those algorithms
which {\it provably} solve a given problem, and have a fast (i.e.\
quickly computable) time bound. Neither the programs themselves,
nor the proofs need to be known in advance. Under these
constraints we construct the asymptotically fastest algorithm
save a factor of 5 that solves any formally defined problem $f$.

\hfill

{\sc Theorem 1.}
{\sl Let $p^*$ be a given algorithm computing $p^*(x)$ from x, or,
more generally, a specification of a function. Let $p$ be any
algorithm, computing provably the same function as $p^*$ with
computation time provably bounded by the function
$t_p(x)$ for all $x$. $time_{t_p}(x)$ is the
time needed to compute the time bound $t_p(x)$.
Then the algorithm $M_{p^*}$ constructed in Section
\ref{secFast} computes $p^*(x)$ in time
\beqn
  time_{M_{p^*}}(x) \;\leq\;
  5\!\cdot\!t_p(x) +
  d_p\!\cdot\!time_{t_p}(x) +
  c_p
\eeqn
with constants $c_p$ and $d_p$ depending on $p$ but not on
$x$. Neither $p$, $t_p$, nor the proofs need to be
known in advance for the construction of $M_{p^*}(x)$.
} 

\hfill

Known time bounds for practical problems can often be computed
very quickly, i.e. $time_{t_p}(x)/time_p(x)$ often converges very
quickly to zero. Furthermore, from a practical point of view, the
provability restrictions are often rather weak. Hence, we
have constructed for every problem a solution, which is
asymptotically only a factor $5$ slower than the (provably)
fastest algorithm! There is no large multiplicative factor, as in
Levin's algorithm, and the problems are not restricted to inversion
problems. What somewhat spoils the practical applicability of $M_{p^*}$ is
the large additive constant $c_p$, which will be estimated in
Section \ref{secTime}.

An interesting consequence of Theorem 1, discussed in Section
\ref{secAIT}, is that the fastest program that computes a certain
function is also one of the shortest programs that provably
computes this function. Looking for larger programs saves, at most, a finite
number of computation steps, but cannot improve the time order.

In Section \ref{secDiscuss} we elucidate the theorem and the range
of applicability on several examples. In Section \ref{secFast} we
give formal definitions of the expressions {\it time}, {\it
proof}, {\it compute}, etc., which occur in Theorem 1, and define
the fast algorithm $M_{p^*}$. The central idea is to enumerate all
programs $p$ equivalent to $p^*$ by enumerating all proofs. In
Section \ref{secTime} we analyze the algorithm $M_{p^*}$,
especially its computation time, prove Theorem 1, and give upper
bounds for the constants $c_p$ and $d_p$. Subtleties regarding the
underlying machine model are briefly discussed in Section
\ref{secAssume}. In Section \ref{secAIT} we show that the fastest
program computing a certain function is also one of the shortest
programs provably computing this function. For this purpose, we
extend the definition of the Kolmogorov complexity of a string and
define two new natural measures for the complexity of functions
and programs. Section \ref{secGeneral} outlines generalizations of
Theorem 1 to i/o streams and other time-measures. Conclusions are
given in Section \ref{secConc}.

\section{Applicability}\label{secDiscuss}

To illustrate Theorem 1, we consider the problem of multiplying
two $n\times n$ matrices. If $p^*$ is the standard algorithm for
multiplying two matrices\footnote{Instead of interpreting $R$ as
the set of real numbers one might take the field
$I\!\!F_2=\{0,1\}$ to avoid subtleties arising from large numbers.
Arithmetic operations are assumed to need one unit of time.}
$x\!\in\!R^{n\cdot n}\!\times\!R^{n\cdot n}$ of size
$\length(x)\!\sim\!n^2$, then $t_{p^*}(x)\!:=\!2n^3$ upper bounds
the true computation time $time_{p^*}(x)\!=\!n^2(2n-1)$. We know
there exists an algorithm $p'$ for matrix multiplication with
$time_{p'}(x)\leq t_{p'}(x)\!:=c\!\cdot\!n^{2.81}$
\cite{Strassen:69}. The time-bound function (cast to an integer)
can, as in many cases, be computed very fast,
$time_{t_{p'}}(x)=O(log^2 n)$. Hence, using Theorem 1, also
$M_{p^*}$ is fast, $time_{M_{p^*}}(x)=5c\!\cdot\!n^{2.81}+O(log^2
n)$. Of course, $M_{p^*}$ would be of no real use if $p'$ is
already the fastest program, since $p'$ is known and could be used
directly. We do not know however, whether there is an algorithm
$p''$ with $time_{p''}(x)\leq d\!\cdot\!n^2log\,n$, for instance.
But if it does exist, $time_{M_{p^*}}(x)\leq
5d\!\cdot\!n^2log\,n\!+\!O(1)$ for all $x$ is guaranteed. The
matrix multiplication example has been chosen for specific
reasons. First, it is not an inversion or optimization problem,
hence unsuitable for Levin search. Second, although matrix
multiplication is a very important and time-consuming issue, $p'$
is not used in practice, since $c$ is so large that for all
practically occurring $n$, the cubic algorithm is faster. The same
is true for $c_p$ and $d_p$, but we must admit that although $c$ is
large, the bounds we obtain for $c_p$ and $d_p$ are tremendous. On
the other hand, even Levin search, which has a tremendous
multiplicative factor, has been successfully applied
\cite{Schmidhuber:97nn,Schmidhuber:97bias}, when handled with
care. The same should hold for Theorem 1, as will be discussed.
We avoid the $O()$ notation as far as
possible, as it can be severely misleading (e.g.\ $10^{42}=O(1)$).

An obvious time bound for $p$ is the actual computation time
itself. An obvious algorithm to compute $time_p(x)$ is to count
the number of steps needed for computing $p(x)$. Hence, inserting
$t_p\!=\!time_p$ into Theorem 1 and using
$time_{time_p}(x)\!\leq\!time_p(x)$, we see that the computation
time of $M_{p^*}$ is optimal within a multiplicative constant
$(d_p+5)$ and an additive constant $c_p$. The result is weaker than
the one in Theorem 1, but no assumption on the computability of
time bounds had to be made.

When do we trust a fast algorithm to solve a problem? At least for
well specified problems, like satisfiability, solving a
combinatoric puzzle, computing the digits of $\pi$, ..., we
usually invent algorithms, prove that they solve the problem and
in many cases also can prove good and fast time bounds. In these
cases, the provability assumptions in Theorem 1 are no real
restriction. The same holds for approximate algorithms which
guarantee a precision $\eps$ within a known time bound (many
numerical algorithms are of this kind). For exact/approximate
programs provably computing/converging to the right answer (e.g.\
traveling salesman problem, and also many numerical programs), but
for which no good, and easy to compute time bound exists, $M_{p*}$
is only optimal apart from a huge constant factor $5+d_p$ in time,
as discussed above. A precursor of algorithm $M_{p^*}$ for this
case, in a special setting can be found in \cite{Hutter:00f}. For
poorly specified problems, Theorem 1 does not help at all.

\section{The Fast Algorithm}\label{secFast}

The idea of the algorithm $M_{p^*}$ is to enumerate proofs of
increasing length in some formal axiomatic system. If a proof
actually proves that $p$ and $p^*$ are functionally equivalent and
$p$ has time bound $t_p$, add $(p,t_p)$ to a list L. The program
$p$ in $L$ with the currently smallest time bound $t_p(x)$ is
executed. By construction, the result $p(x)$ is identical to
$p^*(x)$. The trick to achieve the time bound stated in Theorem
1, is to schedule everything in a proper way, in order not to lose
too much performance by computing slow $p$'s and $t_p$'s before
{\it the} $p$ has been found.

To avoid confusion, we formally define $p$ and $t_p$ to be binary
strings. That is, $p$ is neither a program nor a function, but can
be informally interpreted as such. A formal definition of the
interpretations of $p$ is given below. We say ``p computes
function f'', when a universal reference Turing machine $U$ on
input $(p,x)$ computes $f(x)$ for all $x$. This is denoted by
$U(p,x)\!=\!f(x)$. To be able to talk about proofs, we need a
formal logic system
$(\forall,\lambda,y_i,c_i,f_i,R_i,\rightarrow,\wedge,=,...)$, and
axioms, and inference rules. A proof is a sequence of formulas,
where each formula is either an axiom or inferred from previous
formulas in the sequence by applying the inference rules.
We only need to know that {\it provability}, {\it Turing Machines}, and
{\it computation time} can be formalized:
\begin{enumerate}\parskip=0ex\parsep=0ex\itemsep=0ex
\item The set of (correct) proofs is enumerable.
\item A term $u$ can be defined such that the formula
$[\forall y\!:\!u(p,y)\!=\!u(p^*,y)]$ is true if and
only if $U(p,x)\!=\!U(p^*,x)$ for all $x$, i.e. if
$p$ and $p^*$ describe the same function.
\item A term $tm$ can be defined such that the formula
$[tm(p,x)\!=\!n]$ is true if, and only if the computation time of
$U$ on $(p,x)$ is $n$, i.e.\ if $n\!=\!time_p(x)$.
\end{enumerate}
We say that $p$ is provably equivalent to $p^*$ if the formula $[\forall
y\!:\!u(p,y)\!=\!u(p^*,y)]$ can be proved.

$M_{p^*}$ starts three algorithms $A$, $B$, and $C$ running in
parallel.

\pagebreak[0]
\begin{samepage}\begin{enumerate}\parskip=0ex\parsep=0ex\itemsep=0ex
\item[] {$\nq\nq$\bf Algorithm $M_{p^*}(x)$}
\item[] Initialize the shared variables $L:=\{\},\quad$
$t_{fast}:=\infty,\quad$ $p_{fast}:=p^*$.
\item[] Start algorithms $A$, $B$, and $C$ in parallel with
10\%, 10\% and 80\% \\
computational resources, respectively. \\
That is, $C$ performs 8 steps when $A$ and $B$ perform 1 step each.
\end{enumerate}\end{samepage}

\pagebreak[0]
\begin{samepage}\begin{enumerate}\parskip=0ex\parsep=0ex\itemsep=0ex
\item[] {$\nq\nq$\bf Algorithm $A$}
\item[] $\nq${\tt for} $i$:=1,2,3,... {\tt do}
\item[] pick the $i^{th}$ proof in the list of all proofs and\\
isolate the last formula in the proof.
\item[] {\tt if} this formula is equal to
$[\forall y\!:\!u(p,y)\!=u(p^*,y)\wedge u(t,y)\geq tm(p,y)]$ \\
for some strings $p$ and $t$, \\
{\tt then} add $(p,t)$ to $L$.
\item[] $\nq${\tt next} $i$
\end{enumerate}\end{samepage}

\pagebreak[0]
\begin{samepage}\begin{enumerate}\parskip=0ex\parsep=0ex\itemsep=0ex
\item[] {$\nq\nq$\bf Algorithm $B$}
\item[] $\nq${\tt for} all $(p,t)\!\in\!L$
\item[] run $U$ on all $(t,x)$ in parallel for all $t$ with relative
computational resources $2^{-\length(p)-\length(t)}$.
\item[] {\tt if} $U$ halts for some $t$ and $U(t,x)\!<\!t_{fast}$,\\
{\tt then} $t_{fast}:=U(t,x)$ and $p_{fast}:=p$.
\item[] $\nq${\tt continue $(p,t)$}
\end{enumerate}\end{samepage}

\pagebreak[0]
\begin{samepage}\begin{enumerate}\parskip=0ex\parsep=0ex\itemsep=0ex
\item[] {$\nq\nq$\bf Algorithm $C$}
\item[] $\nq${\tt for} k:=1,2,4,8,16,32,... {\tt do}
\item[] pick the currently fastest program $p:=p_{fast}$
        with time bound $t_{fast}$.
\item[] run $U$ on $(p,x)$ for $k$ steps.
\item[] {\tt if} $U$ halts in less than $k$ steps,
\item[] {\tt then} print result $U(p,x)$ and abort
        computation of $A$, $B$ and $C$.
\item[] $\nq${\tt continue $k$.}
\end{enumerate}\end{samepage}

Note that $A$ and $B$ only terminate when aborted by $C$. The
discussion of the algorithm(s) in the following sections clarifies
details and proves Theorem 1.

\section{Time Analysis}\label{secTime}
Henceforth we return to the convenient abbreviations
$p(x)\!:=\!U(p,x)$ and $t_p(x)\!:=\!U(t_p,x)$. Let $p'$ be some
fixed algorithm that is provably equivalent to $p^*$, with
computation time $time_{p'}$ provably bounded by $t_{p'}$. Let
$\length(proof(p'))$ be the length of the binary coding of the,
for instance, shortest proof. {\it Computation time} always refers
to true overall computation time, whereas {\it computation steps}
refer to instruction steps. $steps=\alpha\!\cdot\!time$, if a
percentage $\alpha$ of computation time is assigned to an
algorithm.\\[-1.5ex]

\noindent A) To write down (not to invent!) a proof requires
$O(\length(proof))$ steps. To check whether the sequence of
formulas constitutes a valid proof requires $O(\length(proof)^2)$
steps. There are less than
$2^{l+1}$ proofs of length $\leq\!l$. Algorithm $A$ receives
$\alpha\!=\!10\%$ of relative computation time. Hence, for a proof
of $(p',t_{p'})$ to occur, and for $(p',t_{p'})$ to be added to
$L$, needs, at most, time $T_A\leq{1\over 10\%}
\!\cdot\!2^{\length(proof(p'))+1}\!\cdot\!O(\length(proof(p')^2)$.
Note that the same program $p$ can and will be accompanied by
different time bounds $t_p$, for instance $(p,time_p)$ will occur.\\[-1.5ex]

\noindent B) The time assignment of algorithm $B$ to the $t_p$'s
only works if the Kraft inequality $\sum_{(p,t_p)\in L}
2^{-\length(p)-\length(t_p)}\leq 1$ is satisfied \cite{Kraft:49}.
This can be ensured by using prefix free (e.g.\ Shannon-Fano)
codes \cite{Shannon:48,Li:97}. The number of steps to calculate
$t_{p'}(x)$ is, by definition, $time_{t_{p'}}(x)$. The relative
computation time $\alpha$ available for computing $t_{p'}(x)$ is
$10\%\!\cdot\!2^{-\length(p')-\length(t_{p'})}$. Hence,
$t_{p'}(x)$ is computed and $t_{fast}\!\leq\!t_{p'}(x)$ is checked
after time $T_B\leq T_A +
10\!\cdot\!2^{\length(p')+\length(t_{p'})} \!\cdot\!
time_{t_{p'}}(x)$. We have to add $T_A$, since $B$ has to wait, in
the worst case, time $T_A$ before it can start executing
$t_{p'}(x)$.\\[-1.5ex]

\noindent C) If algorithm $C$ halts, its construction guarantees that
the output is correct. In the following, we show that $C$ always
halts, and give a bound for the computation time.\\[-3.7ex]
\begin{enumerate}\parskip=0ex\parsep=0ex\itemsep=1ex
\item[\it i)] Assume that
algorithm $C$ stops before $B$ performed the check $t_{p'}(x)<t_{fast}$,
because a different $p$ already computed $p(x)$.
In this case $T_C\leq T_B$.
\item[\it ii)] Assume that $k=k_0$ in
$C$ when $B$ performs the check $t_{p'}(x)<t_{fast}$.
Running-time $T_B$ has passed until this point, hence
$k_0\leq 80\%\!\cdot\!T_B$ . Furthermore, assume that $C$ halts in
period $k_0$ because the program (different from $p'$) executed in
this period computes the result. In this case, $T_C\leq {1\over
80\%}2k_0\leq 2T_B$.
\item[\it iii)] If $C$ does not halt in period $k_0$
but $2k_0\!\geq t_{fast}$, then $p'(x)$ has enough time to compute
the solution in the next period $k=2k_0$, since $time_{p'}(x)\leq
t_{fast}\leq 4k_0-2k_0$. Hence $T_C\leq{1\over 80\%}4k_0\leq 4T_B$.
\item[\it iv)] Finally, if $2k_0\!<t_{fast}$ we ``wait''
for the period $k\!>\!k_0$ with $\odt k\!\leq\!t_{fast}\!<\!k$. In
this period $k$, either $p'(x)$, or an even faster algorithm,
which has in the meantime been constructed by A and B, will be
computed. In any case, the $2k-k>t_{fast}$ steps are sufficient to
compute the answer. We have ${80\%}\!\cdot\!T_C\leq 2k\leq
4t_{fast}\leq 4t_{p'}(x)$.
\end{enumerate}
The maximum of the cases {\it(i)} to {\it(iv)} bounds the
computation time of $C$ and, hence, of $M_{p^*}$ by
$$
  time_{M_{p^*}}(x) = T_C \;\leq\;
  \max\{4T_B,5t_{p}(x)\} \;\leq\;
  4T_B+5t_{p}(x) \;\leq\;
$$ $$
  \;\leq\; 5\!\cdot\!t_p(x) + d_p\!\cdot\!time_{t_p}(x) + c_p
$$ $$
  d_p=40\!\cdot\!2^{\length(p)+\length(t_p)},\quad
  c_p=40\!\cdot\!2^{\length(proof(p))+1}\!\cdot\!O(\length(proof(p)^2)
$$
where we have dropped the prime from $p$. We have also suppressed the
dependency of $c_p$ and $d_p$ on $p^*$ ($proof(p)$ depends on
$p^*$ too), since we considered $p^*$ to be a fixed given
algorithm.

\section{Assumptions on the Machine Model}\label{secAssume}
In the time analysis above we have assumed that program simulation
with abort possibility and scheduling parallel algorithms can
be performed in real-time, i.e.\ without losing performance.
Parallel computation can be avoided by sequentially performing all
operations for a limited time and then restarting all computations
in a next cycle with double the time and so on. This will increase
the computation time of $A$ and $B$ (but not of $C$!) by, at most,
a factor of $4$. Note that we use the same universal Turing
machine $U$ with the same underlying Turing machine model (number
of heads, symbols, ...) for measuring computation time for all
programs (strings) $p$, including $M_{p^*}$. This prevents us from
applying the linear speedup theorem (which is cheating
somewhat anyway), but allows the possibility of designing a $U$
which allows real-time simulation with abort possibility. Small
additive ``patching'' constants can be absorbed in the $O()$
notation of $c_p$. Details will be given elsewhere.

\section{Algorithmic Complexity}\label{secAIT}
Data compression is a very important issue in computer science.
Saving space or channel capacity are obvious applications. A less
obvious (but not far fetched) application is that of inductive inference
in various forms (hypothesis testing, forecasting, classification,
...). A free interpretation of Occam's razor is that the shortest
theory consistent with past data is the most likely to be correct.
This has been put into a rigorous scheme by \cite{Solomonoff:64}
and proved to be optimal in \cite{Solomonoff:78,Hutter:99}.
Kolmogorov Complexity is a universal notion of the information
content of a string \cite{Kolmogorov:65, Chaitin:66, Zvonkin:70}.
It is defined as the length of the shortest program computing
string $x$.
\beqn
  K_U(x) \;:=\; \min_p\{\length(p):U(p)=x\} \;=\; K(x) + O(1)
\eeqn
where $U$ is some universal Turing Machine. It can be shown
that $K_U(x)$ varies, at most, by an additive constant independent
of $x$ by varying the machine $U$. Hence, {\it the} Kolmogorov
Complexity $K(x)$ is universal in the sense that it is uniquely
defined up to an additive constant. $K(x)$ can be approximated from
above (is co-enumerable), but not finitely computable.
See \cite{Li:97} for an
excellent introduction to Kolmogorov Complexity and \cite{Li:00}
for a review of Kolmogorov inspired prediction schemes.

Recently, Schmidhuber \cite{Schmidhuber:01} has generalized
Kolmogorov complexity in various ways to the limits of
computability and beyond. In the following, we also need a
generalization, but of a different kind.
We need a short description of a function, rather than a string.
The following definition of the complexity of a function $f$
\beqn
  K'(f) := \min_p\{\length(p):U(p,x)=f(x)\,\forall x\}
\eeqn
seems natural, but suffers from not even being approximable.
There exists no algorithm converging to $K'(f)$, because it
is undecidable, whether a program $p$ is the shortest program equivalent to a
function $f$. This is similar to the case of the
fastest program. This is obvious if $f$ is an abstract function.
But even if we have a formal specification or program $p^*$ of
$f$, $K'(p^*)$ is not approximable.
Using $K(p^*)$ is not a
suitable alternative, since $K(p^*)$ might be considerably longer
than $K'(p^*)$, as in the former case all information contained
in $p^*$ will be kept -- even that which is functionally
irrelevant (e.g.\ dead code). An alternative is to restrict ourselves to
provably equivalent programs. The length of the shortest one is
\beqn
  K''(p^*) \;:=\;
  \min_p\{\length(p): \mbox{a proof of }
  [\forall y\!:\! u(p,y)=u(p^*,y)] \mbox{ exists} \}
\eeqn
It can be approximated from above, since the set of all programs
provably equivalent to $p^*$ is enumerable.

Having obtained, after some time, a very short description $p'$ of
$p^*$ for some purpose (e.g.\ for defining a prior probability for
some inductive inference scheme), it is usually also necessary to
obtain values for some arguments. We are now concerned with the
computation time of $p'$. Could we get slower and slower
algorithms by compressing $p^*$ more and more? Interestingly this
is not the case. Inventing complex (long) programs is {\it not}
necessary to construct asymptotically fast algorithms, under the
stated provability assumptions, in contrast to Blum's Theorem
\cite{Blum:67,Blum:71}. The following theorem roughly says that {\it there
is a single program, which is the fastest {\rm and} the shortest
program}.

\hfill

{\sc Theorem 2.}
{\sl Let $p^*$ be a given algorithm or
formal specification of a function.
There exists a program $\tilde p$, provably equivalent
to $p^*$, for which the following holds
\beqn
\begin{array}{rl@{\;\leq\;}l}
  i)   & \length(\tilde p)
       & K''(p^*) + O(1) \\[1ex]
  ii)  & time_{\tilde p}(x)
       & 5\!\cdot\!t_p(x) + d_p\!\cdot\!time_{t_p}(x) + c_p
\end{array}
\eeqn
where $p$ is any program provably equivalent to $p^*$ with
computation time provably less than $t_p(x)$.
The constants $c_p$ and $d_p$ depend on $p$ but not on $x$.
} 

\hfill

To prove the theorem, we just insert the shortest algorithm $p'$
provably equivalent to $p^*$ into $M$, that is $\tilde p:=M_{p'}$.
As only $O(1)$ instructions are needed to build $M_{p'}$ from
$p'$, $M_{p'}$ has size $\length(p')\!+\!O(1)=K''(p^*)\!+\!O(1)$.
The computation time of $M_{p'}$ is the same as of $M_{p^*}$
apart from ``slightly'' different constants.

\section{Generalizations}\label{secGeneral}

If $p^*$ has to be evaluated repeatedly, algorithm $A$ can be
modified to remember its current state and continue operation for
the next input ($A$ is independent of $x$!). The large offset time
$c_p$ is only needed on the first call.

$M_{p^*}$ can be modified to handle i/o streams, definable by a
Turing machine with monotone input and output tapes (and
bidirectional working tapes) receiving an input stream and
producing an output stream. The currently read prefix of the input
stream is $x$. $time_p(x)$ is the time used for reading $x$.
$M_{p^*}$ caches the input and output streams, so that algorithm
$C$ can repeatedly read/write the streams for each new $p$. The
true input/output tapes are used, when needing/producing a new
symbol . Algorithm $B$ is reset after $1,2,4,8,...$ steps (not
after reading the next symbol of x!) to appropriately take into
account increased prefixes $x$. Algorithms $A$ just continues. The
bound of Theorem 1 holds for this case too, with slightly increased
$d_p$.

The construction above also works if time is measured in terms of
the current output rather than the current input $x$. This measure
is, for example, used for the time-complexity
of calculating the $n^{th}$ digit of a computable real (e.g.\
$\pi$), where there is no input, but only an output stream.

\section{Summary \& Conclusions}\label{secConc}
We presented an algorithm $M_{p^*}$, which accelerates the
computation of a program $p^*$. The central idea was to enumerate
all programs $p$ equivalent to $p^*$ by enumerating all proofs.
Under certain constraints, $M_{p^*}$ is the asymptotically fastest
algorithm for computing $p^*$ apart from a factor 5 in computation
time. Blum's Theorem shows that the provability constraints are
essential. We have shown that the conditions on Theorem 1 are
often satisfied for practical problems, but not always, however.
For complex approximation problems, for instance, where no good
and fast time bound exists, $M_{p^*}$ is still optimal, but in
this case, only apart from a large multiplicative factor. We
briefly outlined how $M_{p^*}$ can be modified to handle i/o
streams and other time-measures. An interesting consequence of
Theorem 1 was that the fastest program computing a certain
function is also one of the shortest programs provably computing
this function. Looking for larger programs saves, at most, a
finite number of computation steps, but cannot improve the time
order. To quantify this statement, we extended the definition of
Kolmogorov complexity and defined two new natural measures for the
complexity of a function. The large constants $c_p$ and $d_p$ seem
to spoil a direct implementation of $M_{p'}$. On the other hand,
Levin search has been successfully applied to solve rather
difficult machine learning problems
\cite{Schmidhuber:97nn,Schmidhuber:97bias}, even though it suffers
from a large multiplicative factor of similar origin. The use of
more elaborate theorem-provers, rather than brute force
enumeration of all proofs, could lead to smaller constants and
bring $M_p^*$ closer to practical applications, possibly
restricted to subclasses of problems. A more fascinating (and more
speculative) way may be the utilization of so called transparent
or holographic proofs \cite{Babai:91}. Under certain circumstances
they allow an exponential speed up for checking proofs. This would
reduce the constants $c_p$ and $d_p$ to their logarithm, which is
a small value. I would like to conclude with a general question.
Will the ultimative search for asymptotically fastest programs
typically lead to fast or slow programs for arguments of practical
size? Levin search, matrix multiplication and the algorithm
$M_{p^*}$ seem to support the latter, but this might be due to our
inability to do better.

\paragraph{Acknowledgements:}
Thanks to Monaldo Mastrolilli and J{\"u}rgen Schmidhuber for
enlightening discussions and for proof-reading drafts.

\bibliographystyle{alpha}

\end{document}